\documentclass{elsart}

\usepackage{natbib}
\usepackage{graphicx, amssymb}

\usepackage{amssymb}

\begin{document}

\begin{frontmatter}




\title{Heavy meson production at a low-energy photon collider.}


\author{Stephen Asztalos}
\ead{asztalos1\@llnl.gov}
\address{Lawrence Livermore National Laboratory \\ 7500 East Avenue, Livermore, Ca 94551}

\begin{abstract}
  A low-energy $\gamma\gamma$ collider has been discussed in the
  context of a testbed for a $\gamma\gamma$ interaction region at the
  Next Linear Collider (NLC).  We consider the production of heavy
  mesons at such a testbed using Compton-backscattered photons and
  demonstrate that their production rivals or exceeds those by BELLE,
  BABAR or LEP where they are produced indirectly via virtual
  $\gamma\gamma$ luminosities.
\end{abstract}


\end{frontmatter}

\setcounter{table}{0}


\date{\today}

\maketitle

\section{Introduction}
Quantum Chromodynamics (QCD) is a gauged theory of the strong
interactions.  Exquisitely predictive in the perturbative high-energy
regime, at present a description of the physical states is beyond the
reach of the theory. Meson spectroscopy allows a more complete
understanding of QCD in this non-perturbative realm.

Heavy meson spectroscopy provides unique insight into the underlying
theory without the additional complications associated with the the
lighter (and more relativistic) mesons. Heavy mesons have been studied
at $e^{+}e^{-}$ colliders for nearly 30 years, however, only selective
states compatible with the quantum numbers of the $e^{+}e^{-}$ beams
are produced directly: other states are populated via subsequent
hadronic or electromagnetic decays.  Often, as is the case with the
$\eta_b$(nS) mesons, the relevant branching ratios are very small.
Photon interactions provides a mechanism by which some of these
inaccessible mesons are produced directly. In a way that is
complementary to $e^{+}e^{-}$ interactions, $\gamma\gamma$
interactions produce particles having final states with $C$ of +1,
where $C$ is the charge conjugation number.  

Participating photons can be either virtual or real.  Virtual photons
are an irreducible description of charged particle beams: all
e$^+$e$^-$ colliders generate virtual $\gamma\gamma$ luminosity - in
the case of PEP-II or KeK-II the virtual luminosity can rival the
e$^+$e$^-$ luminosity.  An alternate mechanism for producing
$\gamma\gamma$ luminosity is with real photons. Typically, low energy
photons are made to Compton-backscatter off of an energetic charged
particle beam to produce highly boosted photons.  Opposing photon
beams are kinematically focused and interact to produce real
$\gamma\gamma$ luminosity. Although this production mechanism requires
the added complication of introducing a laser and associated optics
into the interaction region, the substantial benefits include control
over the $\gamma\gamma$ luminosity profile, polarization and
magnitude.  Technology advances in optics and lasers have progressed
to the point that one could construct a low energy $\gamma\gamma$
testbed as a proof-of-principle for the NLC, where it would be
possible to singly produce the Higgs boson (whereas it must be
produced in pairs at e$^+$e$^-$ colliders).  The potential this offers
for doing precision Higgs studies is a major driving force behind the
impetus for a $\gamma\gamma$ option at the NLC.  For the purposes of
this study, however, we restrict our attention to photons with a
center of mass energy of 60 GeV, which is sufficient for an
engineering demonstration.  One possible site for the testbed is the
Stanford Linear Collider, where polarized electron beams meet opposing
positrons beams at center-of-mass energies up to the mass of the
Z$_0$. In this proceeding we demonstrate that with suitable choice of
laser and optics parameters the resulting $\gamma\gamma$ luminosity
can be made to exceed the virtual $\gamma\gamma$ luminosity from
e$^+$e$^-$ colliders. This luminosity can thus serve as the basis for
a physics program at the testbed facility, e.g., heavy quarkonia
studies. In subsequent sections we compare and contrast real and
virtual $\gamma\gamma$ luminosities and discuss the implications for
heavy quark meson spectroscopy.
\section{Real photon luminosity}\label{sec:2}
\begin{figure}[!b]
  \begin{center}
    \resizebox{!}{90mm}{\includegraphics[angle=90,totalheight=0.8\textheight,clip]{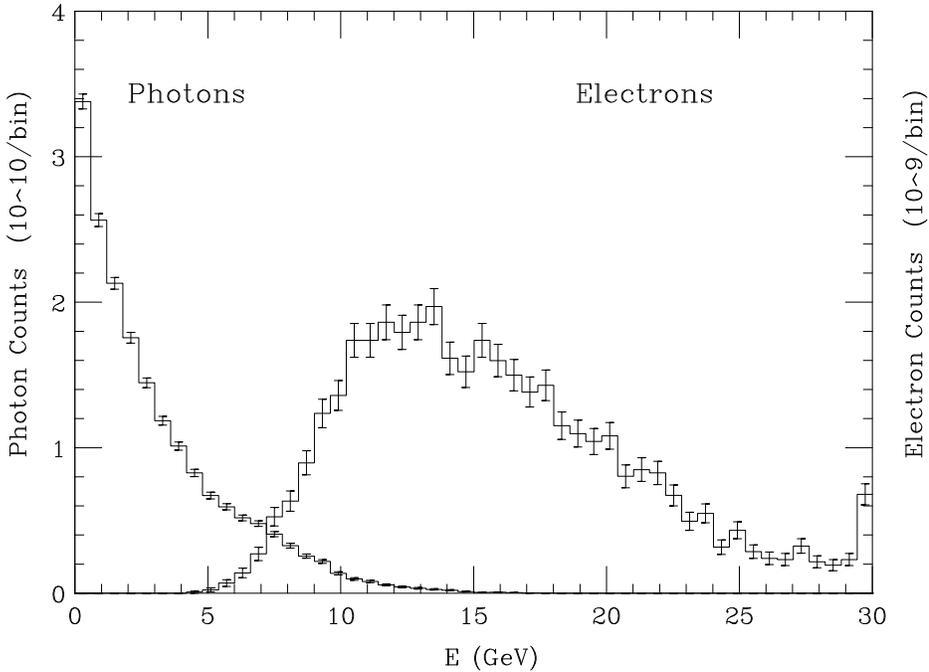}}
  \caption{A snapshot of the right-going photons and electrons at the 
    conversion point.  The ratio of backscattered photons to initial
    electrons is approximately 4:1. Note the different axis scalings
    for photon and electrons.}\label{fig:fig1}
  \end{center}
\end{figure}

Depending on the energy transfer scale the structure of an otherwise
featureless photon can be resolved, resulting in an irreducible hadron
and leptonic background that would complicate precision Higgs
measurements. Hence, a primary goal of a $\gamma\gamma$ testbed would
be to measure the $\gamma\gamma$ luminosity and compare the result
with simulation. In this study Compton-backscattered luminosity was
simulated with CAIN 2.1e \cite{CAIN} using the laser and beam
parameters listed in Table \ref{tab:t1}, where the electron and laser
polarizations refer to circular polarization.
\renewcommand\baselinestretch{0.7}
\begin{table}[t!] 
  \begin{centering} {\footnotesize
      \begin{tabular}{|c|c||c|c|}
        \hline
        \multicolumn{2}{|c||}{Beam}  & \multicolumn{2}{c|}{Laser}\\\hline
        parameter & value            & parameter & value \\\hline\hline
        beam energy & 30 GeV         & frequency & 120 Hz\\\hline
        charge & 4$\times$10$^{10}$         & pulse energy & 2.0 J\\\hline
        $\sigma_x$ & 1.47 $\mu$m     & $\lambda$ & 1.05 $\mu$m\\\hline
        $\sigma_y$ & 0.0552 $\mu$m   & laser polarization &$\pm$ 1\\\hline
        $\sigma_z$ & 0.1 mm         & &\\\hline
        $\beta_x$ & 8.0 mm            & &\\\hline
        $\beta_y$ & 0.1 mm          & &\\\hline 
        $e^-$ polarization & 0.8          & &\\\hline 
        $e^+$ polarization & 0      & &\\\hline 
      \end{tabular} }
    \caption{\footnotesize Beam and laser parameters used to characterize the
      real photon luminosity at LINX.}\label{tab:t1}
  \end{centering}
\end{table}
\renewcommand\baselinestretch{1.0} Beam parameters in this table
reflect NLC-like expectations about achievable beam sizes, while the
laser parameters were chosen to maximize the Compton-backscattering
rate. (Throughout this proceeding we refer to a hypothetical
accelerator whose beam parameters match those of Table \ref{tab:t1} as
LINX.) Figure \ref{fig:fig1} shows the resulting simulated
Compton-backscattered electron and resultant photon spectra at the
LINX conversion point.
With nearly 2$\times$10$^{10}$ photons per electron the conversion is
very efficient: more than 98\% of the electrons undergo
Compton-scattering. The preponderance of photons also causes
significant degradation of the single-interaction photon and electron
Compton edges around 10 and 20 GeV, respectively.  This degradation is
due to multiple Compton interactions as the electron beam traverses
the laser pulse. This picture is quite distinct from the sharp Compton
edges seen in simulations (not shown) where the energy of the laser
pulse is considerably lower and single interactions dominate.

\begin{figure}[h]
  \begin{center}
  \resizebox{!}{90mm}{\includegraphics[angle=90]{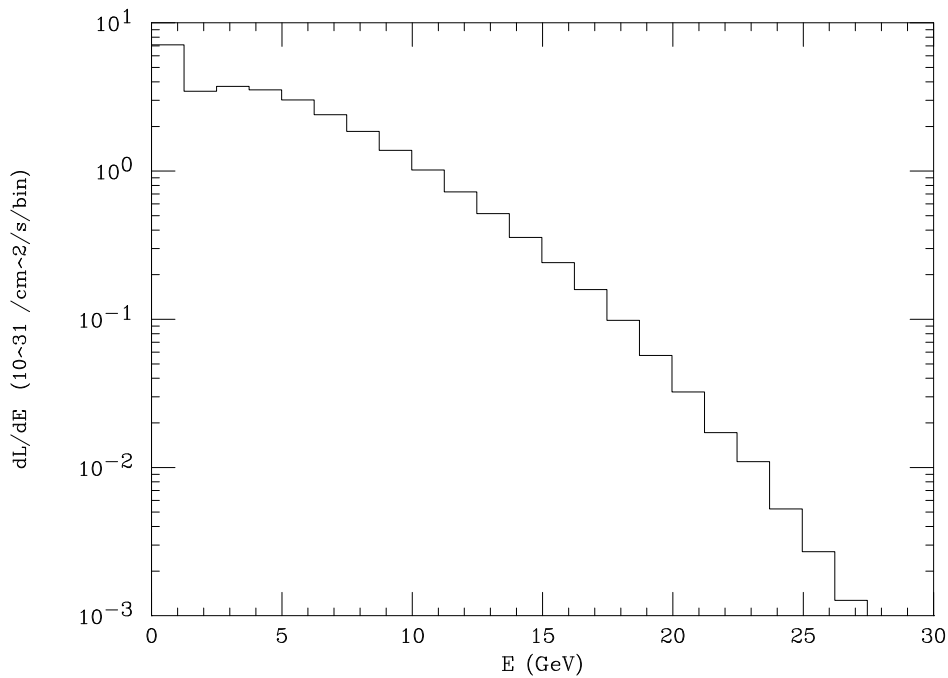}}
  \caption{$\gamma\gamma$ luminosity for the beam and laser
    parameters in Table \protect{\ref{tab:t1}}.  The total integrated
    (instantaneous) luminosity is
    2.971$\times$10$^{32}$cm$^{-2}$s$^{-1}$}\label{fig:fig_2}
  \end{center}
\end{figure}

$\gamma\gamma$ luminosity is produced when the left- and right-going
photons traverse the several millimeters from their respective
conversion points to the interaction point. Figure \ref{fig:fig_2}
shows the resulting simulated Compton-backscattered $\gamma\gamma$
luminosity at the interaction point from CAIN.  For this analysis the
luminosity is divided into 50 equally spaced energy bins spanning the
range 0 $\!<\!\!\sqrt{s}\!<\!$ 62.4 GeV.
Multiple Compton interactions give rise to significant luminosity
above 20 GeV, whereas for lower laser energies (where single
interactions dominate) the luminosity is essentially zero for energies
above twice the Compton edge.  The area under the curve in Fig.
\ref{fig:fig_2}, or the instantaneous $\gamma\gamma$ luminosity, is
2.971$\times$10$^{32}$cm$^{-2}$s$^{-1}$.  By way of comparison, the
maximum instantaneous e$^+$e$^-$ luminosity at LEP never exceeded
1.0$\times$10$^{32}$cm$^{-2}$s$^{-1}$ over its 10+ years of operation.
\section{Virtual photon luminosity}\label{sec:3}
To assess the relative merits of a real photon collider we have
considered three colliders whose large e$^+$e$^-$ luminosities would
suggest a large virtual $\gamma\gamma$ luminosity: PEP-II \cite{PEP},
KEK-B \cite{KEK} and LEP \cite{OPAL}.  PEP-II and KEK-B are asymmetric
$e^{+}e^{-}$ colliders running at a center of mass energy of 10.58
GeV, while LEP was a symmetric $e^{+}e^{-}$ machine that scanned over
the energy range 91.2 $\!<\!\!$ $\sqrt{s}$ $\!\lesssim\!$ 206 GeV
until decommissioned in late 2000 to make way for the Large Hadron
Collider (LHC).

The classical treatment of virtual photons exploits the equivalence
between the fields of energetic charged particles and pulses of
electromagnetic radiation.  In the equivalent photon approximation
(EPA) a pulse of electromagnetic radiation is transformed into a
frequency spectrum of virtual photons.  The interaction
of virtual photons from colliding charged particle beams gives rise to
virtual photon luminosity.  The Weizsacker Williams approximation to
the EPA restricts itself to transverse modes of the virtual photon
while integrating over the (unobserved) momentum transfer between the
electron and the virtual photon to yield  \cite{ww:34}
\begin{equation}
\frac{dN}{dz}=\frac{\alpha}{2\pi z}\left[\left( 1 - z + \frac{z^2}{2} \right)
    \!\log (\frac{q^2_{max}}{q^2_{min}}) -  2m^2_e\left( \frac{1}{q^2_{min}}  - 
       \frac{1}{q^2_{max}} \right)\!z^2\right]\label{eq:1}
\end{equation}
where $z$ is defined as E$_\gamma$/E$_{beam}$, $m_e$ the electron (or
positron) mass, $q^2_{min}$ and $q^2_{max}$ are the minimum and
maximum photon energies, respectively, and $dN/dz$ is the number of
photons in an energy interval $dz$.  Both $q^2_{min}$ and
$q^2_{max}$ depend on $z$ in the following manner
\begin{eqnarray}
  q^2_{min} & = & \frac{m_e^2z^2}{1-z} \\
  q^2_{max} & = & ( 1 - z )E_{beam}^22\left(1-\cos\theta_{max}\right)
  \label{eq:3} 
\end{eqnarray}
The dependence of $q^2_{max}$ on $\theta_{max}$ in Eq. \ref{eq:3}
merits comment.  As $\theta_{max}$ increases the opportunity for
detecting electrons diminishes due to limited detector acceptance, but
it is precisely these unobserved electrons that contribute to Eq.
\ref{eq:1}. 
\begin{table}[!b]
  \begin{center} {\footnotesize
      \begin{tabular}{|c|c|c|c|c|}
        \hline
       & BABAR & BELLE  & OPAL & LINX\\\hline
       Energy (GeV) & 3.1/9.0 & 3.5/8.0  & 91-206/91-206 & 30/30\\\hline
       $\theta_{max}$ (\footnotesize mrad) & 523.6/295.6  & 715.6/401.4 & 33/33 & 36/36 \\\hline
      \end{tabular} }
  \end{center}
  \caption{\footnotesize $e^{+}/e^{-}$ beam energies  and 
    forward/backward acceptances for the BABAR, BELLE, OPAL and LINX detectors. }\label{tab:tab2}
\end{table}
The appearance of $\theta_{max}$ the above expression requires that
specific detector configurations be taken into account in the WW
analyses. BABAR \cite{BABAR} and BELLE \cite{BELLE} detectors surround
the interaction regions at PEP-II and KEK-B, respectively.  We take
the OPAL \cite{OPAL} detector as the prototypical LEP detector, while
the acceptances of the LINX detector were assumed to be those of the
Stanford Linear Detector \cite{SLD}.  Table \ref{tab:tab2} lists the
beam energies and forward and backward acceptances seen by the four
detectors under consideration. The forward and backward acceptances of
BABAR and BELLE are unequal due to the asymmetric beam energies.

In order to compare real and virtual $\gamma\gamma$ luminosities
yearly e$^+$e$^-$ luminosity records for BABAR, BELLE and OPAL were
compiled.  Table \ref{tab:tab3} summarizes the peak and integrated
$e^{+}e^{-}$ luminosities seen by each detector since initiation of
their respective operations.  \renewcommand\baselinestretch{0.7}
\begin{table}[t!]
  \begin{center} {\footnotesize
      \begin{tabular}{|c|c|c|c|c|c|c|}
        \hline
        & \multicolumn{2}{c|}{BABAR} & \multicolumn{2}{c|}{BELLE} &
        \multicolumn{2}{c|}{OPAL} \\\cline{2-7}
        year & \multicolumn{1}{c|}{peak\footnotemark}
        & \multicolumn{1}{c|}{int\footnote{In units of fb$^{-1}$}} &
        \multicolumn{1}{c|}{peak} & \multicolumn{1}{c|}{int} & \multicolumn{1}{c|}{peak} &
        \multicolumn{1}{c|}{int} \\\hline\hline
        1990 &- &-   &-    &-  & 0.0110\footnotemark  & 0.0121          \\
        1991 &- &-   &-    &-  & 0.0110\footnotemark[3] & 0.0189            \\
        1992 &- &-   &-    &-  & 0.0110\footnotemark[3] & 0.0286               \\
        1993 &- &-   &-    &-  & 0.0190  & 0.0400            \\
        1994 &- &-   &-    &-  & 0.0231  & 0.0645             \\
        1995 &- &-   &-    &-  & 0.0341  & 0.0461               \\
        1996 &- &-   &-    &-  & 0.0356  & 0.0247          \\
        1997 &- &-   &-    &-  & 0.0570  & 0.0734            \\
        1998 &- &-   &-    &-  & 0.0999  & 0.1997               \\
        1999 &1    &1.62  &2.05\footnotemark  &0.287    & 0.1069 & 0.2537            \\
        2000 &2    &23.76 &2.05   &10.940   & 0.0675  & 0.2331             \\
        2001 &4.4  &40.05 &5.171  &36.257   & - &    -              \\
        2002 &5    &31.32 &8.256  &54.181   & - &    -                \\
        2003 &7.5  &56.71 &11.305 &77.232   & - &    -                \\
        2004 &7.5\footnotemark[5]  &4.63   &11.305\footnotemark[5] &6.827   & - &    -       \\
        \hline\hline Total  &  -  &  158.09 & & 185.724 &  - &  0.9948                 \\
        \hline
      \end{tabular} }
  \end{center}
  \caption{\footnotesize Instantaneous and integrated $e^{+}e^{-}$ luminosities 
    as a function of year. For BABAR and BELLE the 
    2004 luminosities are through 01/04. }\label{tab:tab3}
\end{table}
\addtocounter{footnote}{-3} \footnotetext{In units of
  nb$^{-1}$s$^{-1}$} \stepcounter{footnote}\footnotetext{In units of
  fb$^{-1}$} \stepcounter{footnote}\footnotetext{LEP peak luminosities
  assumed for 1990-1992.}  \stepcounter{footnote}\footnotetext{BELLE
  peak luminosity not available for 1999.}
\stepcounter{footnote}\footnotetext{BELLE and BABAR peak luminosities
  for 2004 taken from 2003.}
\section{Results}\label{sec:sec4}
\begin{figure}[h!]
  \begin{center}
    \resizebox{!}{80mm}{\includegraphics[angle=0]{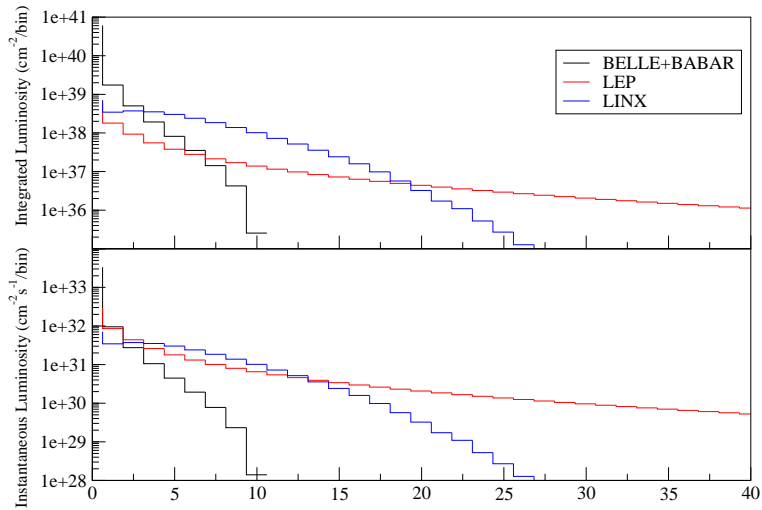}}
    \caption{The instantaneous (lower) 
      and integrated (upper) virtual Weizsacker-Williams luminosities
      seen by BELLE+BABAR and LEP compared to the real luminosity seen
      at LINX. All luminosities are in 1.3 GeV bins.}\label{fig:fig4}
  \end{center}
\end{figure}

To simulate Weizsacker Williams luminosity the physics event generator
PANDORA 2.1 \cite{PAN} was used. Since Eq. \ref{eq:1} increases
without limit as z$\rightarrow$0, it was necessary to introduce a
low-energy cutoff to regularize the result.  An energy cutoff of 1 MeV
was employed in all subsequent analyses.  The choice of cutoff
obviously affects the integrated virtual luminosity, but has no
discernible affect on the virtual photon spectra above the energy
range of interest ($\sim$ 3 GeV), nor on the luminosity integrated
cross sections discussed below.  Virtual $\gamma\gamma$ luminosity was
calculated for the PEP-II, KEK-B and LEP beams using Pandora in two
steps.  First, a random number corresponding to the energy of a
virtual in one beam was generated.  Associated with this photon energy
is a specific value of $dN/dz$ from the Weizsacker Williams
approximation.  This same procedure was repeated for the opposing
beam.  The product of the $dN/dz$ from each opposing beam gives the
relative virtual $\gamma\gamma$ luminosity per unit $e^{+}e^{-}$
luminosity, which for PEP-II was determined to be 0.19. Instantaneous
virtual $\gamma\gamma$ luminosities were derived by multiplying the
relative virtual $\gamma\gamma$ luminosities by the instantaneous
$e^{+}e^{-}$ luminosities from Table \ref{tab:tab3}.  Although the
table contains yearly peak luminosities (when available), for our
purpose the peak luminosity with the greatest magnitude was used.  A
comparison with the LINX instantaneous $\gamma\gamma$ luminosity is
shown in the lower graph in Fig.  \ref{fig:fig4}. In both graphs in
this figure and in the analyses described below the BELLE and BABAR
luminosities have been summed together, while the LEP luminosity is
that seen by the OPAL detector, but multiplied by four to account for
the four LEP detectors.
The characteristic rise inherent in the Weizsacker Williams
description of virtual photons is seen in the BELLE+BABAR and LEP
plots, whereas the rise in the LINX luminosity at low energies stems
instead from bremsstrahlung and multiple scattering.  Using the
relative $\gamma\gamma$ luminosities above and the peak luminosity
data from Table \ref{tab:tab3}, one arrives at peak instantaneous
luminosities of 3.479$\times$10$^{33}$cm$^{-2}$s$^{-1}$ and
3.121$\times$10$^{32}$cm$^{-2}$s$^{-1}$ for BABAR+BELLE and LEP,
respectively. By comparison, the LINX instantaneous luminosity is
2.971 $\times$10$^{32}$cm$^{-2}$s$^{-1}$.

A similar approach was used in producing the integrated luminosities
in the top graph in Fig.  \ref{fig:fig4}, but here the relative
virtual $\gamma\gamma$ luminosities are multiplied by the yearly
integrated luminosities and summed over relevant years to form a total
integrated luminosity.  For instance, in the case of LEP there are 11
contributions to the sum (years 1990 to 2000) - each year with its own
relative and integrated luminosity to account for the changing beam
energy with time.  A summary of the results for the normalized,
instantaneous and integrated luminosities is contained in Table
\ref{tab:lum_sum}.  \renewcommand\baselinestretch{0.7}
\begin{table}[t!]
  \begin{center} {\footnotesize
      \begin{tabular}{|l|c|c|c|}
        \hline
               & BABAR+BELLE & LEP & LINX \\\hline
        norm   & 0.19        & 0.52-0.70   & 1.0  \\\hline
        inst (10$^{32}$cm$^{-2}$s$^{-1}$)  & 34.79 & 3.121 & 2.971 \\\hline
        intr (10$^{39}$cm$^{-2}$s)        & 63.86 & 2.60  & 2.971\\\hline
      \end{tabular} }
  \end{center}
  \caption{\footnotesize  Normalized, instantaneous and integrated luminosities
  from BABAR+BELLE, LEP and LINX.  One Snowmass year of operation (10$^7$ s) was
  assumed for the LINX integrated luminosity.}\label{tab:lum_sum}
\end{table}
Taken together, Table \ref{tab:lum_sum} and the top graph in Fig.
\ref{fig:fig4} is one of the our main findings: although the
integrated virtual $\gamma\gamma$ luminosity from BELLE+BABAR greatly
exceeds the integrated real $\gamma\gamma$ luminosity expected from
LINX, due to the sharply decreasing nature of the Weizsacker-Williams
spectrum this dominance disappears beyond 3.0 GeV.  As will be seen in
the next section, since much of the physics of interest lies above
this energy, real $\gamma\gamma$ luminosity offers a very viable approach
to heavy meson spectroscopy.
\section{Physics Opportunities}
One goal of heavy quark spectroscopy is to test Quantum Chromodynamics
predictions of the masses, widths and quantum numbers of these mesons
by comparing them with their experimental values. Production of
pseudoscalar mesons is a particularly attractive physics goal for a
low-energy $\gamma\gamma$ collider since the mesons quantum numbers
J$^{CP}$ = 0$^{+-}$ preclude them from being produced directly at an
$e^{+}e^{-}$ collider.  Although the lowest radial excitations of the
charmonium system has been thoroughly mapped out (with the exception
of the singlet $h$(1p) state), spectroscopy of the second radial
excitation has numerous holes. Even where there have been recent
measurements, as in the case of the pseudoscalar $\eta_c$(2S),
disagreement persists as to the excitation (mass).  Conflict between
experiment and theory on the order of 50 MeV in mass of the
$\eta_c$(2S) is the source of some consternation in the theoretical
community \cite{etac}. Enhanced statistics would further allow for the
determination of unknown branching ratios. The situation is more
severe in the bottom system: none of the pseudoscalar bottom mesons
have been definitively observed \cite{etab}.  Bottom pseudoscalar
mesons apparently are more difficult to produce at $e^{+}e^{-}$
colliders: their heavier heavier masses imply a more non-relativistic
wave function.  This, in turn, enhances the selection rule which in
the non-relativistic limit forbids $E$(1) transitions between the
$\Upsilon$(nS)$\rightarrow$$\eta_b$(nS).

Using the integrated luminosities described in the previous section we
have studied the production of pseudoscalar and other suitable heavy
charm and bottom mesons at a low-energy $\gamma\gamma$ testbed and
have compared these results to what can be achieved through virtual
$\gamma\gamma$ luminosity.  Meson simulations were again carried out
using the PANDORA event generator. Meson production in PANDORA is
modeled using the Breit-Wigner approach; necessary modifications were
made to include the widths and masses of the various mesons that are
summarized in Table \ref{tab:t4}.
\begin{table}[t!]
  \begin{center} {\footnotesize
      \begin{tabular}{|c||c|c|c|}
        \hline
        meson         & mass    & $\Gamma_{tot}$ &  $\Gamma_{\gamma\gamma}$      \\\hline\hline
        $\eta_c$(1S)  & 2.979   & 0.0161         &  7.4$\times$10$^{-6}$   \\\hline
        $\eta_c$(2S)  & 3.654   & 0.0161         &  7.4$\times$10$^{-6}$   \\\hline
        $\chi_{c0}$   & 3.415   & 0.0107         &  2.6$\times$10$^{-6}$   \\\hline
        $\eta_b$(1S)  & 9.3     & 0.014          &  5.3$\times$10$^{-7}$   \\\hline
      \end{tabular} }
  \end{center}
  \caption{\footnotesize Masses and widths of the mesons in GeV.}\label{tab:t4}
\end{table}
To model meson production at LINX CAIN 2.1 files were imported into
PANDORA. These CAIN luminosity files have 10000 entries representing
the partitioning of the energy range 0 $\!<\!$ $\sqrt{s}$ $\!<$ 62.4
GeV into 50 equally spaced energy bins of 1.248 GeV each for
E$_{\gamma1}$ and E$_{\gamma2}$.  There are four sets of 2500 entries
corresponding to the four helicity combinations possible when two
photons interact.  Each of the 10000 entries contains E$_{\gamma1}$
and E$_{\gamma2}$, luminosity and overall helicity.  PANDORA
normalizes the luminosity so that the sum over all 10000 entries is
1.0.  Meson production at BABAR+BELLE and LEP was modeled using the
luminosity class described in Section \ref{sec:3}.

PANDORA can be made to output luminosity integrated cross sections
for the process of interest, defined as
\begin{equation}
  \overline{\sigma} \equiv \int \frac{\partial L}{\partial E} \sigma dE
\end{equation}
In practice, the integral is replaced by a finite sum sampled from a
uniformly distributed E$_{\gamma1}$ and E$_{\gamma2}$ plane.  The
utility of the luminosity integrated cross section is that when
multiplied by the total integrated luminosity the result is the number
of events for the process under consideration.
\renewcommand\baselinestretch{0.7}
\begin{table}[b!]
  \begin{center} {\footnotesize
      \begin{tabular}{|c|c|c|c|c|c|c|}
        \hline
        & \multicolumn{2}{c|}{BABAR+BELLE} & \multicolumn{2}{c|}{LEP} &
        \multicolumn{2}{c|}{LINX} \\\cline{2-7}
        meson & \multicolumn{1}{c|}{$\overline{\sigma}$}
        & \multicolumn{1}{c|}{events} 
        & \multicolumn{1}{c|}{$\overline{\sigma}$}
        & \multicolumn{1}{c|}{events}
        & \multicolumn{1}{c|}{$\overline{\sigma}$}
        & \multicolumn{1}{c|}{events} \\\hline
        $\eta_c$(1S)     & 3.19e+4 & 1.08e+7  & 1.72e+6  & 1.98e+6 & 3.34e+6  & 9.91e+6      \\\hline
        $\eta_c$(2S)     & 1.25e+4 & 4.25e+6  & 8.43e+5  & 9.74e+5 & 1.97e+6  & 5.84e+6      \\\hline
        $\chi_{c0}$      & 5.23e+3 & 1.78e+6  & 3.32e+5  & 3.83e+5 & 7.66e+5  & 2.28e+6      \\\hline
        $\eta_b$(1S)     & 3.05e-1 & 1.09e+2  & 3.25e+2  & 3.89e+2 & 1.07e+3  & 3.17e+3      \\\hline
      \end{tabular} }
  \end{center}
  \caption{\footnotesize The luminosity integrated cross sections and
  events produced for the three machine scenarios considered.}\label{tab:events}
\end{table}
\renewcommand\baselinestretch{1.}  Table \ref{tab:events} lists the
luminosity integrated cross sections and total number of events
produced for each meson in the previous table.  Note that unlike the
conventional definition of cross section, the luminosity integrated
cross section depends on details of the environment in which the meson
is produced: it is lowest for BABAR+BELLE, for which the integrated
luminosity is greatest. From this same table it is seen that number of
$\eta_c$(1S) events produced at BABAR+BELLE and LINX are approximately
equal.  Figure \ref{fig:fig4} provides a consistency check on this
result. In the upper plot in that figure the integrated luminosities
of BABAR+BELLE and that of LINX are seen to be approximately equal at
an energy corresponding to the mass of the $\eta_c$(1S).  Beyond this
crossover point, a greater number of heavier mesons are produced at
the LINX facility than at BELLE+BABAR or LEP.  For the purposes of
bottom spectroscopy the LINX scenario is clearly superior, with the
potential for producing an order of magnitude or more $\eta_b$(1S)
mesons than either BELLE+BABAR or LEP.

\renewcommand\baselinestretch{1.}  In order to simulate experimental
reconstruction of these mesons at a $\gamma\gamma$ testbed it is
necessary to specify the decay mode, as well as potential backgrounds.
Decay of heavy mesons into p$\rm\bar{p}$ pairs was chosen due to the
simplicity of the event reconstruction, the relatively large branching
ratios and knowledge about the competing background process.  The
dominant background $\gamma\gamma\rightarrow$p$\rm\bar{p}$ arises when
both photons are resolved into hadrons.  This process has been
well-studied for insight into the photon structure function and
resolution of the ongoing debate concerning diquark versus independent
quark models of the hadron.  Theoretical \cite{diqt} and experimental
\cite{diqe} efforts have gone into determining the cross section for this
process.  In particular, the latest BELLE results describe a
differential cross section (in nb/sr) for this process of the form
\begin{equation}
\frac{d\sigma}{dEd(\cos\theta)}=1.15\!\times\!10^6\sigma^{15}\left(\frac{1+\cos^2\theta}
  {1-\cos^2\theta}\right)\label{eq:back}
\end{equation}
for $|\cos(\theta)| \le$ 0.6 and $\sqrt s \ge$ 2.75 GeV. This result
is consistent with the diquark model, in contrast to the independent
quark model predicts an isotropic angular distribution.  The
normalization factor that appears in Eq. \ref{eq:back} was derived
directly from BELLE data. A separate Pandora class was written to
generate the $\gamma\gamma\rightarrow$p$\rm\bar{p}$ background; when
convolved with the LINX luminosity described above a luminosity
integrated cross section subject to the above kinematic limits of
0.0058 nb is obtained, which corresponds to 17337 events over a
Snowmass year of operation.  Similarly, to compute the number
reconstructed meson events it was necessary to multiply the raw number
of meson events from the last row of Table \ref{tab:events} by the
branching ratio for decay into p$\rm\bar{p}$ pairs.  Table
\ref{tab:final} summarizes the number of events expected from this
procedure.  
\renewcommand\baselinestretch{.7}
\begin{table}[h!]
  \begin{center} {\footnotesize
      \begin{tabular}{|c||c|c|c|c|}
        \hline
        meson            & BR(p$\rm\bar{p}$)        &  p$\rm\bar{p}$   &Reconstructed p$\rm\bar{p}$ events \\\hline\hline
        $\eta_c$(1S)     & 0.0012                   &  1.19e+4         &  7.09e+3   \\\hline
        $\eta_c$(2S)     & 0.00049                  &  2.80e+3         &  1.69e+3   \\\hline
        $\chi_{c0}$      & 0.00024                  &  5.47e+2         &  3.32e+2   \\\hline
        $\eta_b$(1S)     & 0.0012                   &  4.00e+0         &  3.00e+0   \\\hline\hline
        $\gamma\gamma\rightarrow$p$\rm\bar{p}$&  -  &  1.73e+4         &  1.73e+4   \\\hline\hline
        $\rm$ Total      &  -                       &  3.26e+4         &  2.64e+4    \\\hline
      \end{tabular} }
  \end{center}
  \caption{\footnotesize Branching ratios into p$\rm\bar{p}$ final states and total
    number of signal and background events expected for one Snowmass year of running
    at a $\gamma\gamma$ testbed.}\label{tab:final}
\end{table}
\renewcommand\baselinestretch{1.0}The Phythia interface to Pandora
allows events to be written in StdHeP format \cite{stdhep}, which in
turn are read in by the LCDROOT \cite{lcdroot} - a detector simulation
and analysis framework based on ROOT \cite{root}. In this analysis the
p$\rm\bar{p}$ final states and background were propagated through a
small version of a hypothetical NLC detector whose electromagnetic and
hadronic calorimeters reside in 5 T magnetic field.  Details of the
detector \cite{sld} are not crucial to the results.  Passage of
p$\rm\bar{p}$ through the detector simulation results in momenta
smearing.  In the hadronic calorimeter the momenta is inferred.  For
this analysis only inclusive decays to p$\rm\bar{p}$ are considered,
thus all particles reaching the hadronic calorimeters were assumed to
have the mass of a proton. With this information and the total energy
measured in the electromagnetic calorimeter the four vector is
constructed and the invariant mass is calculated.  
\begin{figure}[h!]
  \begin{center}
    \resizebox{!}{70mm}{\includegraphics[angle=0]{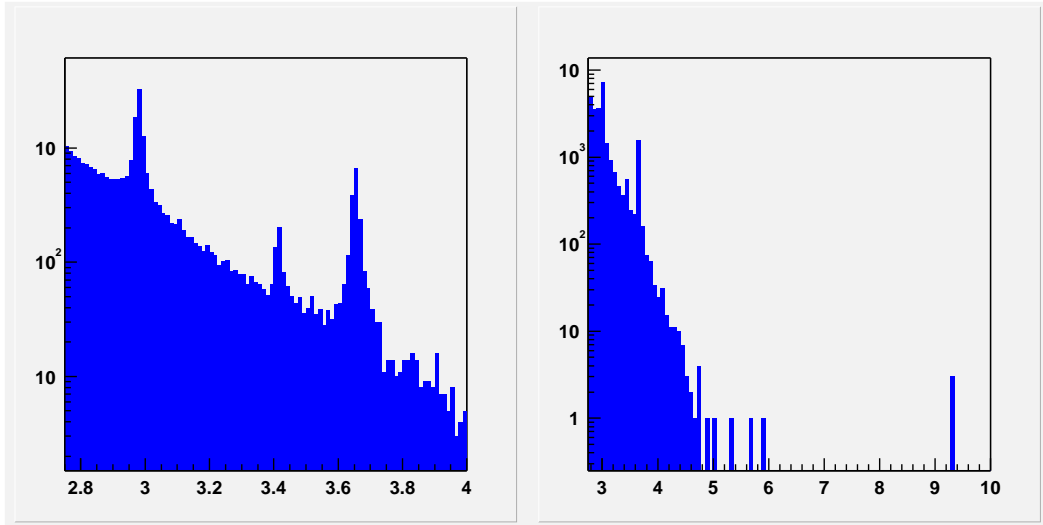}}
    \caption{Reconstructed meson and background events expected from 
      a Snowmass year of operation at LINX. Lower and upper plots are
      identical except for the scaling of the axes.  Cuts of
      $|\cos(\theta)| \le$ 0.6 and $\sqrt s \ge$ 2.75 GeV were applied
      to all data.}\label{fig:events}
  \end{center}
\end{figure}
Figure \ref{fig:events} is a reconstruction of the invariant mass of
events leaving a track in the hadronic calorimeter subject to the
restriction that $|\cos(\theta)| \le$ 0.6 and $\sqrt s \ge$ 2.75 GeV.
The total number of reconstructed events (26,494) found in the last
column of \ref{tab:final} is significantly fewer than the 32,527
StdHeP events input into LCDROOT.  This is an artifact stemming from
the fact that kinematic cuts were not imposed on the meson events
within PANDORA but rather inside LCDROOT.  For the isotropically
decaying mesons the fraction lost is consistent with the restricted
phase space imposed by the angular cuts.  The meson events clearly
stand out above the p$\rm\bar{p}$ background, although additional
analysis is needed to determine whether it is possible to separate out
$\chi_{c0}$ and $\eta_c$(2S) events. A thorough description of these
results, including a complete treatment of the non-isotropic decay of
the $\chi_{c2}$, will appear elsewhere.
\section{Conclusions}
A $\gamma\gamma$ interaction region has been proposed for the NLC
using Compton-backscattered photons. Technical challenges integrating
the optical, laser and mechanical subsystems suggest the desirability
of a low-energy testbed.  In this paper we address one aspect of a
physics program that could accompany such a test bed - heavy meson
production - and compare it with similar meson production expected
where virtual photons are the means of production.  We find that after
a single year of operation the number of charmed mesons seen at a LINX
facility equals or exceeds the number generated at either BABAR+BELLE
or LEP to-date. LINX is especially dominant when producing bottom
mesons, offering the exciting possibility of discovering $\eta_b$(nS)
mesons before the LHC begins delivering data toward the end of this
decade.
\section{Acknowledgments}
The authors would like to thank Prof. Robert Nisius for helpful
insight into the virtual photon methodology.


\begin{thebibliography}{00}
\bibitem{CAIN} http://www-acc-theory.kek.jp/members/cain/
\bibitem{PEP} http://www.slac.stanford.edu/accel/pepii/home.html
\bibitem{KEK} http://belle.kek.jp/
\bibitem{OPAL} http://opal.web.cern.ch/Opal/
\bibitem{ww:34}
K. F. von Weizsacker, Z. Physik 88, 612 (1934), \\E. J. Williams, Phys. Rev. 45, (1934) 729. 
\bibitem{BABAR} http://www.slac.stanford.edu/BFROOT/
\bibitem{BELLE} http://belle.kek.jp/
\bibitem{SLD} http://www-sld.slac.stanford.edu/sldwww/sld.html
\bibitem{PAN} http://www-sldnt.slac.stanford.edu/nld/new/Docs/Generators/PANDORA.htm
\bibitem{etac} DELPHI Collaboration, J. Abdallah, et al, Eur. Phys.J.
  C31 (2003) 481
\bibitem{etab} ALEPH Collaboration, Phys.Lett. B530, (2002) 56
\bibitem{diqt} V. M. Budnev et al., Phys. Rep. 15, (1975) 181
\bibitem{diqe} Ch.-Ch. Kuo et al., Nucl. Phys.B (Proc. Suppl.) 126, (2004) 313
\bibitem{stdhep} http://www-cpd.fnal.gov/psm/stdhep/
\bibitem{lcdroot} http://www-sldnt.slac.stanford.edu/nld/New/Docs/LCD\_Root/root.htm
\bibitem{root} http://root.cern.ch/
\bibitem{sld} http://wwwal.kuicr.kyoto-u.ac.jp/www/accelerator/sspm/Iwasaki.pdf
  

\end{thebibliography}
\end{document}